\documentstyle[amssymb,amsmath]{elsart}
\newcommand{\kk}{\rangle \!\rangle}
\newcommand{\bb}{\langle \!\langle}
\newcommand{\ket}[1]{ | \, #1  \rangle}
\newcommand{\bra}[1]{ \langle #1 \,  |}
\newcommand{\kett}[1]{ |  #1  \rangle\!\rangle}
\newcommand{\braa}[1]{ \langle \!\langle #1   |}
\newcommand{\proja}[1]{\ket{#1}_a \,{}_a\bra{#1}}
\newcommand{\projb}[1]{\ket{#1}_b \,{}_b\bra{#1}}
\begin{document}
\begin{frontmatter}
\title{Unitary realizations of the ideal phase measurement}
\author{F. Buscemi, G. M. D'Ariano, and  M. F. Sacchi}
\address{Quantum
Optics \& Information Group\\ Universit\`a degli Studi di Pavia and INFM Unit\`a di
Pavia\\ via A. Bassi 6, I-27100 Pavia, Italy} 
\begin{abstract}
We explicitly construct a large class of unitary transformations that
allow to perform the ideal estimation of the phase-shift on a
single-mode radiation field.  The ideal  phase distribution is
obtained by heterodyne detection on two radiation modes after the
interaction.
\end{abstract}
\end{frontmatter}
The quantum estimation of an unknown phase shift---the so called
quantum phase measurement---is the essential problem of high sensitive
interferometry, and has received much attention in quantum optics
\cite{rev1}. For a single-mode electromagnetic field, the measurement
cannot be achieved exactly, even in principle, due to the lack of a
unique self-adjoint operator \cite{london}. In fact, the absence of a
proper self-adjoint operator is mainly due to the semi-boundedness of
the spectrum of the number operator \cite{shsh,ban}, which is
canonically conjugated to the phase in the sense of a
Fourier-transform pair \cite{shap}.

This observation opened the route for an exact phase measurement in
terms of two-mode fields, where a phase-difference operator becomes
conjugated to an unbounded number-difference operator \cite{luis}. In
fact, a concrete experimental setup using unconventional heterodyne
detection has been suggested \cite{unc} for this kind of measurement.
However in the single-mode case, no feasible scheme that can provide
the optimal phase measurement has been devised yet.

The most general and concrete approach to the problem of the phase
measurement is quantum estimation theory \cite{helstrom}, a framework
that has become popular only in the last ten years in the field of
quantum information.  Quantum estimation theory provides a more
general description of quantum statistics in terms of POVM's (positive
operator-valued measures) and gives the theoretical definition of an
optimized phase measurement. The most powerful method for deriving the
optimal phase measurement was given by Holevo \cite{Holevo} in the
covariant case.  In this way the optimal POVM for phase estimation has
been derived for a single-mode field.  More generally, the problem of
estimating the phase shift has been addressed in Ref. \cite{chiara}
for any degenerate shift operator with discrete spectrum, either
bounded, bounded from below, or unbounded, extending the Holevo
method for the covariant estimation problem.

\par As already stated, quantum estimation theory provides the optimal
POVM for the phase measurement. This writes in terms of projectors on
Susskind-Glogower states \cite{ssg}
\begin{eqnarray}
d\mu (\varphi )=\frac {d\varphi}{2\pi }|e^{i\varphi } \rangle \langle
e^{i\varphi } |\;,\label{sg}
\end{eqnarray}
where $|e^{i\varphi }\rangle =
\sum_{n=0}^{\infty} e^{i\varphi n}|n \rangle $. 
Notice that the states $|e^{i\varphi }\rangle $ are not normalizable, neither
orthogonal, however they provide a resolution of the identity, and
thus guarantee the completeness of the POVM, namely
\begin{eqnarray}
\int _{0}^{2\pi }d\mu (\varphi )=I
\end{eqnarray}
For a system in state $\rho$, the POVM in Eq. (\ref{sg}) gives the
ideal phase distribution $p(\varphi )$ according to Born's rule 
\begin{eqnarray}
p(\varphi )=\hbox{Tr}[d\mu (\varphi )\,\rho]= 
\frac {d\varphi}{2\pi }\langle e^{i\varphi } |\rho |
e^{i\varphi } \rangle \;.\label{pfi}
\end{eqnarray}
\par In this Letter we will explicitly construct some unitary
transformations 
that allows to perform the ideal phase measurement described by the
POVM in Eq. (\ref{sg}). First, we will introduce an isometry $\tilde V$
which enlarges the Hilbert space of the system (say ${\cal H}_a$ for
mode $a$) to the tensor product ${\cal H}_a \otimes {\cal H}_b$ for
two modes $a$ and $b$. Then, we will prove that the exact 
measurement of the complex photocurrent $Z=a-b^\dag$ provides through
its marginal distribution the ideal probability density $p(\varphi )$
of Eq. (\ref{pfi}).  Finally, we will construct a large class of
unitary operators on ${\cal H}_a \otimes {\cal H}_b \otimes {\cal
  H}_c$, where ${\cal H}_c$ denotes the Hilbert space of an ancillary
arbitrary system, such that the isometry $\tilde V$ is realized with
unit probability. 
\par  We start by introducing the eigenstates of the heterodyne
photocurrent $Z=a-b^\dag$ \cite{yuen,shwa,unc}
\begin{eqnarray}
Z |D(z) \kk _{ab} =z |D(z) \kk _{ab}\;,
\end{eqnarray}
where $D(z)=\exp(za^\dag -z^*a)$ denotes the displacement operator. 
Here and in the following we use the notation \cite{pla} for bipartite
pure states on ${\cal H}_a \otimes {\cal H}_b $
\begin{eqnarray}
|A{\rangle\!\rangle} _{ab}=\sum_{n,m=0}^{\infty } 
A_{nm}|n\rangle _a\otimes|m\rangle _b \equiv A\otimes
I_b |I{\rangle\!\rangle}_{ab}
 \equiv I_a \otimes A^\tau |I{\rangle\!\rangle}_{ab}
\;, 
\end{eqnarray} 
where $A^\tau $ denotes the transposed operator with respect to some
pre-chosen orthonormal basis.  The states $|D(z) \kk _{ab}$ are
orthogonal in Dirac sense over the complex plane, namely
\begin{eqnarray}
{}_{ab}\bb D(\alpha )|D(\beta )\kk _{ab}=\pi \delta^{(2)}(\alpha -\beta )
\equiv \pi \delta (\hbox{Re}\ \alpha -\hbox{Re}\ \beta )
\,\delta (\hbox{Im}\ \alpha -\hbox{Im}\ \beta )  \;.
\end{eqnarray}
They also provides a basis for ${\cal H}_a \otimes {\cal H}_b$ as
follows  
\begin{eqnarray}
\int _{\mathbb C}\frac {d^2 z}{\pi } |D(z) \kk _{ab}\, {}_{ab}\bb  D(z)|=I_a\otimes
I_b
\;.\label{pomz}
\end{eqnarray}
The measurement of the complex photocurrent $Z$ can be performed
through unconventional heterodyne detection \cite{shwa} with both the
signal $a$ and the image-band $b$ non-vacuum (in usual heterodyne
detection the image-band mode is in the vacuum, thus providing the
well-known coherent-state POVM).  The measurement of 
$Z$ is also equivalent to two separate homodyne
measurements on modes $\frac {1} {\sqrt 2}(b \pm a)$.  In fact,
consider the $50/50$ beam splitter operator $R=\exp [\frac \pi 4
(a^\dag b-a b^\dag )]$ that realizes the unitary transformation
\begin{eqnarray}
R
\left( \begin{array}{c}a\\ b\end{array}\right)
R^\dag ={\frac {1}{\sqrt 2}}
\left( \begin{array}{lr} 
1 & -1\\ 
1& 1\end{array}\right) 
\ \left( \begin{array}{c}a\\ b\end{array}\right)
\;.\label{matrix}
\end{eqnarray}
Upon denoting with $\ket{x}_a$ and $\ket{y}_b$ the eigenstates of the
quadratures $X_a= (a+a^\dag )/\sqrt 2$ and $Y_b=(ib^\dag -ib)/\sqrt
2$, one has the following identity \cite{unc2}
\begin{eqnarray}
R (\proja{x} \otimes \projb{y}) R^\dag  = 
\kett{D(x+iy)}_{ab}\,{}_{ab} \braa{D(x+iy)}
\;.\label{1}
\end{eqnarray}
Notice also that this kind of measurement is performed in the
teleportation protocol for continuous variable of  Braunstein-Kimble
scheme \cite{tele1,tele2}. 

\par We can now write the isometry $\tilde V$ such that 
the transformation 
\begin{eqnarray}
T(\rho )= \tilde V \rho \tilde V ^\dag \;\label{iso}
\end{eqnarray}
maps the state of the system $\rho \in {\cal H}_a$ to a two-mode state in 
${\cal H}_a \otimes {\cal H}_b$. The operator $\tilde V$ has the form 
\begin{eqnarray}
\tilde V&=&\frac 1 {\sqrt {2\pi}}\int _{\mathbb C}d^2 \alpha \,f(|\alpha |) 
|D(\alpha )\kk _{ab}\, {}_a\langle e^{i\arg\alpha  }|
\;.\label{form}
\end{eqnarray}
By choosing $f(t)$ as an arbitrary function satisfying the condition
\begin{eqnarray}
\int_{0}^{+\infty }dt\,t\,|f(t)|^2 =\frac {1}{\pi }\;,\label{cond}
\end{eqnarray}
it follows that $\tilde V$ is an isometry, namely $\tilde V^\dag
\tilde V = I_a$. 
\par It is easy to check that the
transformation (\ref{iso}) has the following covariance symmetry
\begin{eqnarray}
T(e^{i\theta a^\dag a} \rho e^{-i \theta a^\dag a })=
e^{i\theta a^\dag a} \otimes 
e^{-i \theta b^\dag b } T(\rho )\,e^{-i\theta a^\dag a} \otimes 
e^{i \theta b^\dag b }
\;.
\end{eqnarray}
We can now evaluate the probability density of getting outcome $z \in
\mathbb C$ through the measurement of the photocurrent $Z$. One has 
\begin{eqnarray}
p(z)&=&\frac {1}{\pi}
\hbox{Tr}[\tilde V\rho \tilde 
V^\dag |D(z)\kk _{ab}\, {}_{ab}\bb D(z)|]\nonumber \\&= &
\frac 12 |f(|z|)|^2 {}_a \langle e^{i \varphi }|\rho |e^{i\varphi
}\rangle _a
\;,\qquad
\varphi =\arg z
\end{eqnarray}
From condition in Eq. (\ref{cond}), it follows that 
the marginal distribution on the statistical variable $\varphi =\arg
z$ corresponds to the ideal distribution of the phase for any
input state $\rho $.
\par In the following we explicitly construct some unitary realizations of
the map in Eq. (\ref{iso}). We start by defining the operators in
${\cal L}({\cal H}_a\otimes {\cal H}_b)$ 
\begin{eqnarray}
V =\tilde V (I_a \otimes {}_b\langle \chi |)\;,\qquad 
V ^\dag =(I_a \otimes |\chi \rangle _b) \tilde V^\dag \;,\label{chi}
\end{eqnarray}
where $|\chi \rangle $ is an arbitrary normalized state in ${\cal H
}_b$, and the tensor notation $I_a \otimes {}_b \langle \chi |$
represents a linear operator from ${\cal H}_a \otimes {\cal H}_b$ to
${\cal H}_a $ (the bra ${}_b \langle \chi |$ can be regarded as a linear
functional from ${\cal H}_b$ to $\mathbb C$). Similarly, $I_a \otimes
|\chi \rangle _b$ represents an operator from ${\cal H}_a $ to ${\cal
H}_a \otimes {\cal H}_b$.  Notice that both $VV^\dag $ and $V^\dag V$
are projectors, namely \begin{eqnarray} VV^\dag VV^\dag =VV^\dag
\;,\qquad V^\dag VV^\dag V=V^\dag V\;.
\end{eqnarray}
Upon introducing an arbitrary Hilbert space ${\cal H}_c$ (also finite
dimensional), we construct the
following operator 
\begin{eqnarray}
U=V \otimes WW^\dag 
-V^\dag \otimes W^\dag W 
+(I-  V^\dag  V ) 
\otimes W^\dag + (I- V V^\dag )\otimes W 
\;,\label{11}
\end{eqnarray}
where $W$ is a linear operator in ${\cal H}_c$. Under the conditions 
\begin{eqnarray}
W^2=(W^\dag)^2=0\;, \qquad 
WW^\dag +W^\dag W =I_c\;,\label{condw}
\end{eqnarray}
one can easily check that $WW^\dag $ and $W^\dag W$ are projectors
orthogonal each other, and $U $ is unitary.  Consider now the
transformation of the system prepared in a state $\rho \otimes \sigma
\otimes \mu $ which has been evolved through the unitary $U$ and
traced over the ancillary space ${\cal H}_c$. One has
\begin{eqnarray}
&&\hbox{Tr}_c[U (\rho \otimes \sigma \otimes \mu )U^\dag ] 
 =  V (\rho \otimes \sigma )V^\dag  \hbox{Tr}[WW^\dag \mu ] \nonumber
 \\& & 
+V (\rho \otimes \sigma )(I- VV^\dag )  \hbox{Tr}[W^\dag \mu ]
+V^\dag  (\rho \otimes \sigma )V  \hbox{Tr}[W^\dag W \mu ]
\nonumber \\& & -V^\dag  (\rho \otimes \sigma )(I-V^\dag V)  \hbox{Tr}[W \mu ]
 -(I-V^\dag V) (\rho \otimes \sigma )V
 \hbox{Tr}[W^\dag  \mu ]
\nonumber \\& & 
+(I-V^\dag V) (\rho \otimes \sigma )(I- V^\dag V)  \hbox{Tr}[WW^\dag
 \mu ]
+(I-V V^\dag ) (\rho \otimes \sigma )V^\dag   \hbox{Tr}[W  \mu ]
\nonumber \\& & 
+(I-V V^\dag ) (\rho \otimes \sigma )(I-  V V^\dag )  \hbox{Tr}[W^\dag W
 \mu ]
\;.\label{trc}
\end{eqnarray}
The map in Eq. (\ref{iso}) can then be achieved by the unitary
transformation $U$, by taking 
\begin{eqnarray}
\sigma =| \chi \rangle _b \,{}_b\langle \chi |\;,
\end{eqnarray}
and $\mu $ such that 
\begin{eqnarray}
\hbox{Tr}[WW^\dag \mu ]=1\;,\qquad 
\hbox{Tr}[W \mu ]=\hbox{Tr}[W^\dag \mu ]=0 \;.\label{condmu}
\end{eqnarray}
We summarize the conditions on the measurement scheme: $f(t), |\chi
\rangle , W, \mu $ in Eqs. (\ref{form}), (\ref{chi}), (\ref{11}),
(\ref{trc}) are arbitrary, provided that conditions (\ref{cond}),
(\ref{condw}), (\ref{condmu}) are satisfied.  

\par As an example, consider the case in which the space ${\cal H}_c$
pertains to a radiation mode $c$. One can take
\begin{eqnarray}
W=\sum_{n=0}^\infty |2n \rangle _c \,{}_c \langle 2n+1 |\;,
\qquad  \mu =|0 \rangle _c \,{}_c\langle 0 |
\;,\label{wn}
\end{eqnarray}
thus obtaining 
\begin{eqnarray}
T(\rho )=\hbox{Tr}_c [U (\rho \otimes 
|\chi \rangle _b \, {}_b \langle \chi | \otimes |0 \rangle _c 
\, {}_c \langle 0 |)U^\dag ]\;.\label{ob}
\end{eqnarray}
We notice that the ``pseudo-spin'' operator $W$ in Eq. (\ref{wn}) has
been introduced also in Refs. \cite{pspin1,pspin2} in the context of
Bell's inequalities for continuous variable.  

\par The result in Eq. (\ref{ob}) is similarly obtained for a qubit system in ${\cal
H}_c$, with $W=|0 \rangle _c \,{}_c \langle 1|$.  

\par In conclusion, a large class of unitary realizations of the ideal
phase measurement of a single-mode radiation field has been
presented. These unitary evolutions act on the Hilbert space ${\cal
H}_a\otimes {\cal H}_b\otimes {\cal H}_c$, where ${\cal H}_a$ and
${\cal H}_b$ are referred to radiation modes, and ${\cal H}_c$
pertains to an arbitrary system. By suitably preparing the state of
the systems in ${\cal H}_b$ and ${\cal H}_c$, the ideal 
phase distribution for the input state $\rho \in {\cal H}_a$ is
obtained through heterodyne detection performed after the interaction
on modes $a$ and $b$.

\section*{Acknowledgments}
This work has been sponsored by INFM through the project
PRA-2002-CLON, and by EEC through the ATESIT project IST-2000-29681.

\end{document}